# Aegis:An Advanced LLM-Based Multi-Agent for Intelligent Functional Safety Engineering

**Lu Shi, Bin Qi, Jiarui Luo, Yang Zhang,**
**Zhanzhao Liang**, **Zhaowei Gao**, **Wenke Deng**, **Lin Sun**
{lu.shi, bin.qi, jiarui.luo, yang.zhang2, }@hirain.com
{zhanzhao.liang, zhaowei.gao, wenke.deng, lin.sun}@hirain.com

## Abstract

Functional safety is a critical aspect of automotive engineering, encompassing all phases of a vehicle's lifecycle, including design, development, production, operation, and decommissioning. This domain involves highly knowledge-intensive tasks. This paper introduces Aegis: An Advanced LLM-Based Multi-Agent for Intelligent Functional Safety Engineering. Aegis is specifically designed to support complex functional safety tasks within the automotive sector. It is tailored to perform Hazard Analysis and Risk Assessment (HARA), document Functional Safety Requirements (FSR), and plan test cases for Automatic Emergency Braking (AEB) systems. The most advanced version, Aegis-Max, leverages Retrieval-Augmented Generation (RAG) and reflective mechanisms to enhance its capability in managing complex, knowledge-intensive tasks. Additionally, targeted prompt refinement by professional functional safety practitioners can significantly optimize Aegis's performance in the functional safety domain. This paper demonstrates the potential of Aegis to improve the efficiency and effectiveness of functional safety processes in automotive engineering.

## 1 Introduction

The functional safety requirements cover all activities throughout the vehicle's lifecycle, including design, development, production, operation, and decommissioning (International Organization for Standardization, 2011). According to ISO 26262, functional safety activities for on-road vehicles, compliant with regulations and project experience, are organized according to the V-model, covering all critical activities from the concept phase to the decommissioning phase, as illustrated in Figure 1.

Implementing functional safety requires thorough knowledge of standards like ISO 26262 and IEC 61508, covering safety requirements from analysis to maintenance, and necessitates professional expertise (Nouri and Warmuth, 2021). High-level systems thinking, statistical skills, and deep domain knowledge are essential for identifying hazards and analyzing risks using techniques like Fault Tree Analysis (FTA) and Failure Mode and Effects Analysis (FMEA) (Cristea and Constantinescu, 2017). Defining safety requirements and designing effective safety mechanisms involve interdisciplinary knowledge in hardware design, software development, and safety engineering. Achieving Safety Integrity Level (SIL) requires rigorous verification and validation through extensive testing, including functional verification, software and hardware testing, system integration testing, and validation of Safety of the Intended Functionality (SoV) and Safety of the Intended Use (SoC) (International Organization for Standardization, 2011). Configuration management and change control are crucial for maintaining system safety throughout the product lifecycle, involving tracking and assessing changes to prevent new risks (International Organization for Standardization, 2011). Continuous learning and knowledge updates are essential due to evolving automotive E/E systems and advancements in autonomous driving algorithms (Martin et al., 2016; Chen et al., 2024). These characteristics fully demonstrate that functional safety activi-ties are knowledge-intensive work which refers to tasks that require significant cognitive effort and specialized expertise to complete.

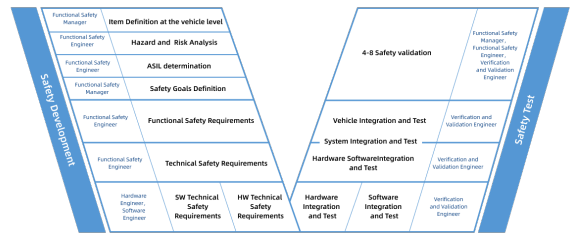

Figure 1: The V-Model of Functional Safety Activities and Roles

Large Language Models (LLMs) are highly appropriate for addressing knowledge-intensive tasks owing to their robust capabilities in knowledge acquisition, storage, and application (AlKhamissi et al., 2022). LLMs have already been used in HARA analysis (Nouri et al., 2024). However, LLMs can sometimes generate inaccurate information, especially when dealing with domain-specific or complex issues (Kandpal et al., 2023). For instance, if an LLM is provided with a functional requirement for Automatic Emergency Braking (AEB) and tasked with conducting a Hazard Analysis and Risk Assessment (HARA) in accordance with UL4600, it may not produce an accurate response if it has not been trained on the UL4600 regulations.

To address such situations, Retrieval-Augmented Generation (RAG) can incorporate external knowledge from databases to solve these domain-specific, knowledge-intensive tasks (Lewis et al., 2020). Additionally, training and fine-tuning LLMs to locate and modify specific knowledge stored within the models can also address information gaps or inaccuracies (De Cao et al., 2021; Yao et al., 2023; Mitchell et al., 2022).

Considering that pre-training large models is a resource-intensive process with high costs, and that fine-tuning still demands substantial computational resources—with costs varying according to task complexity, data volume, and model size (Liu et al., 2023)—we propose using RAG to extend LLM knowledge in the specific domain of functional safety. RAG allows for low-cost integration of new domain knowledge by incorporating both the internal and external functional safety regulations, automotive E/E system requirements, papers verification and validation processes, and other expert knowledge into external databases (Vector Database and File System).

By employing retrieval, generation, and augmentation techniques, RAG supports the entire functional safety lifecycle. This approach not only enhances the LLM's capabilities in functional safety but also ensures that the system remains up-to-date with the latest domain-specific information.

LLMs have the distinct capability of assuming different roles when given specific identity prompts, thereby simulating the social division of labor in the real world. LLM-based multi-agents enhance task performance through social behaviors such as collaboration and competition. These agents can encourage divergent thinking, improve reasoning capabilities, and reduce hallucinations, making them well-suited for handling complex knowledge tasks.

In functional safety activities, as illustrated in Figure 1, various roles such as Functional Safety Manager, V&V Engineer, and others are involved. These roles collaborate to accomplish complex functional safety tasks that span different domains, such as HARA analysis and functional safety validation. By establishing a multi-agent system where each agent focuses on its specific tasks within the functional safety lifecycle, they can collectively achieve the overall functional safety goals through coordinated efforts.

In this paper, we propose Aegis, an LLM-based multi-agent system designed to support functional safety activities. The system is specifically tailored to carry out Hazard Analysis and Risk Assessment (HARA), Functional Safety Requirements (FSR) documentation, and test case planning tasks for an Automatic Emergency Braking (AEB) system. Additionally, it automatically creates associations and mappings between Safety Goals (SG), FSR, and test cases.

In comparison to existing tools like medini analyze® and Vector Informatik, Aegis's key innovation lies in its higher level of automation. While current tools require significant manual input, Aegis introduces a hierarchical multi-agent framework and Retrieval-Augmented Generation (RAG) to dynamically integrate external standards (e.g., ISO 26262, VDA 702), providing real-time compliance updates. This significantly enhances both the automation and precision of complex functional safety tasks.

We designed three versions of Aegis based on the LLM QWEN-MAX which is is a trillion-parameter large-scale language model from Alibaba (Alibaba, 2024):

1. **Aegis-Lite**: Comprising 2 agents: functional safety manager and verification and validation engineer.

2. **Aegis-Pro**: Comprising 3 agents: functional safety manager, verification and validation engineer and functional safety expert.

3. **Aegis-Max**: Comprising 3 agents, enhanced with Retrieval-Augmented Generation (RAG),

and incorporating reflection and critique mechanisms.

We also introduced professional functional safety practitioners to provide few-shot prompts and conducted two rounds of targeted prompt refinement to guide the agents in performing higher-quality functional safety activities.

To evaluate the task outcomes, we established a set of assessment criteria derived from experienced functional safety experts and regulations. Both GPT-4o and seasoned functional safety experts scored and assessed the agents' outputs multiple times.

The findings indicate that Aegis-pro, by adding more agent roles compared to Aegis-Lite, increased the accuracy of HARA analysis and FSR generation while reducing incorrect responses. With improved prompts, the agents provided more accurate answers to detailed queries. Furthermore, the inclusion of RAG and reflection mechanisms in Aegis-max enhanced the comprehensiveness of HARA analysis and the coverage of generated test cases.

## 2 Aegis Design

Aegis-Max aims to automate functional safety activities for AEB requirements. Its primary functions include performing functional safety HARA analysis, developing FSRs, and writing test cases. Aegis-Max integrates multiple roles and components, including the Functional Safety Manager, Functional Safety Expert, and Verification and Validation (V&V) Engineer, each with specific tasks and responsibilities. In Aegis, agents independently perform tasks like hazard analysis or test case planning. Each agent operates autonomously within its role and coordinates with others to achieve common goals, ensuring flexibility and efficiency in handling complex functional safety tasks.

Figure 2 shows the workflow of Aegis-max and the description is below:

**Input** User provides the AEB requirement and poses the question: "Please generate the functional activities with the input requirement {REQUIREMENT}."

The document is divided into smaller chunks with a size of 2000 and an overlap of 10 to avoid issues caused by exceeding the length limitation of QWEN-MAX.

**Aegis-Max** Aegis-Max is a multi-agent system representing a functional safety team.

**Functional Safety Manager** This role encompasses the combined tasks of the Functional Safety Manager and Functional Safety Engineer as defined in Figure 1. For prompt details regarding role definitions, please refer to Appendix BB.1. In smaller functional safety teams, it is common for a single engineer to handle the responsibilities of both roles. Additionally, to reduce communication overhead between agents and improve efficiency (Qian et al., 2023), we have assigned the duties of both roles to the Functional Safety Manager within Aegis-Max. We define that the Functional Safety Manager needs to conduct safety definitions and safety analyses, explicitly stating the need to refer to the VDA 702 Standard in the knowledge base for HARA analysis. In Section 3, Experiments (Prompt) and Evaluation, the results are also described, demonstrating that HARA Analysis yields better outcomes through RAG.

Additionally, by strictly defining the output format of the Functional Safety Manager's results after performing safety analyses like HARA and FTA through few-shot prompts, as detailed in Appendix B.1, we improve the controllability and consistency of the agent's output (Ding et al., 2023).

**Functional Safety Expert** This role encompasses more extensive knowledge and insights related to functional safety, as detailed in Appendix B.1. The role is defined as "more professional than the functional safety manager." In this role, a higher-level review process is also defined, allowing the Expert to critique the Manager's work from a higher dimension and update the safety planning content based on these critiques.

**V&V Engineer** We assigned the role of functional safety verification and validation engineer to the V&V Engineer. This role involves planning tests based on the messages output by the Functional Safety Expert, and producing consistent test case tables according to specific formats. At this stage, we did not provide detailed prompts for generating test cases, such as test case coverage. Instead, by assigning the role to the V&V Engineer, the agent's outputs are expected to align with the role's definition (Park et al., 2023).

**Self-RAG** A reflection RAG for Few-shot prompts. It includes two main roles: Researcher and Revisor. For each functional safety-related role, after experienced functional safety engineers have evaluated the results generated without the reflec-tion process, we detailed the reflection and critique process for each role based on their suggestions. For example, when the V&V engineer conducts a reflection, they need to consider the coverage of the test cases. Detailed content can be found in Appendix B.1.

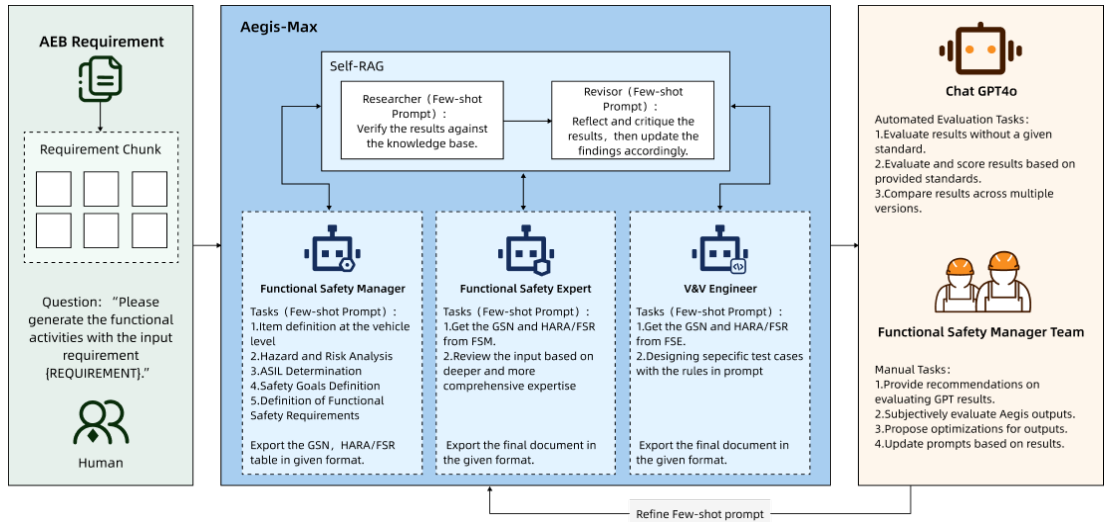

Figure 2: The workflow of Aegis-Max

**Researcher (Few-shot Prompt)** This node functions as a RAG query mechanism, primarily responsible for searching various documents within the knowledge base, including regulatory texts, best practice documents, and functional requirement case studies. Its role is to update the outputs from preceding role nodes while maintaining the original output format. The knowledge base service leverages Alibaba's BAILIAN platform application center. By constructing a knowledge repository on BAILIAN, RAG queries are executed via API calls using QWEN-MAX-based application APIs. The construction and implementation details of RAG itself fall outside the scope of Aegis's discussion.

**Revisor (Few-shot Prompt)** Given that our application scenarios and outputs are well-defined, and we seek more in-depth and accurate responses from Aegis regarding functional safety activities, the Revisor node provides targeted prompts based on the specific roles of the agents. This ensures task clarity and accessibility, reducing the likelihood of hallucinations in complex tasks and keeping the results focused on the core responsibilities of each actor (Khademi, 2023)

**Evaluation and Reflection** We evaluated the outputs generated by Aegis, with GPT-4o and human functional safety engineers scoring and assessing the Functional Safety Requirements (FSR) and test cases.

**Chat GPT-4o** Detailed descriptions of automated evaluation tasks can be found in Chapter 3, "Experiments and Evaluation." Automated evaluations were conducted by GPT-4o using custom evaluation templates designed by experienced functional safety engineers. Additionally, to discuss the impact of RAG and multi-role supervision on knowledge-intensive and complex functional safety tasks, we designed Aegis-Lite Figure 3 and Aegis-Pro Figure 4 for comparative evaluation of the three agent frameworks.

**Functional Safety Manager Team** An experienced team of functional safety managers also scored and assessed the results. Additionally, they provided new suggestions for prompts to improve the accuracy of Aegis's outputs.

**Interaction** Interaction in Aegis is entirely goal-driven, not based on negotiation. Each agent has a defined role, such as generating a HARA report or refining outputs for test cases. Agents work sequentially, sharing and updating outputs based on feedback. This structured, goal-oriented interaction improves accuracy through iterative feedback, enabling efficient management of complex tasks with minimal errors.

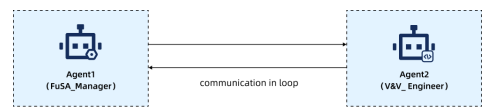

Figure 3: Aegis-Lite: Includes only FuSA_Manager and V&V_Engineer, completing tasks through multi-agent dialogue.

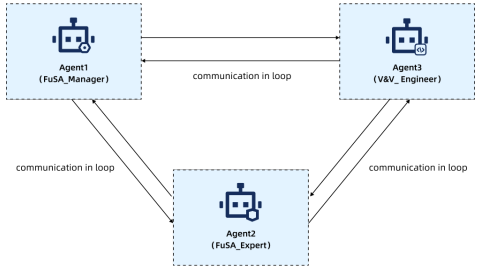

Figure 4: Aegis-Pro: Adds a supervisory node, FuSA_Expert, to complete functional safety activities through mutual dialogue, but does not include RAG.

## 3 Experiments and Evaluation

To evaluate Aegis's performance in executing complex functional safety tasks, we tested and assessed Aegis-Lite, Aegis-Pro, and Aegis-Max.

We conducted two types of evaluations: (1) Human evaluation, and (2) GPT-4o evaluation (Bran et al., 2023). For vehicle functional safety, Aegis provides 20 functional safety requirements and corresponding test cases for the vehicle each time it runs, presenting a comprehensive final solution. This solution is then compared with a single solution generated by the GPT-4o model. To ensure fairness, the GPT-4o was also provided with the relevant knowledge base documents and the same prompts. See Appendix A.1 for details.

### 3.1 Evaluation Criteria

The evaluation criteria were formulated by several professional automotive safety testing experts with over five years of industry experience, based on the "Functional Safety Review and Evaluation Methods" published by the China National Standardization Management Committee(of People's Republic of China, 2023), ISO 26262(International Organization for Standardization, 2011), and their professional experience.

The evaluation criteria is attached in Appendix D.

#### 3.1.1 Experiment Process

We conducted experiments with different prompts and agent frameworks, obtaining a total of seven sets of functional safety requirements and test case results, as shown in the Table 1 below:

For detailed prompt content during the iteration process, refer to Appendices B.1, B.2, and B.3.

The few-shot prompt is present in detail in Appendix A. The difference among the three versions of the prompt is summarized below:

**Initial Prompt** The first version which can induce the FuS_Manager and V&V Engineer can export the FSR and test cases.

**Second Version** Refined based on the initial version. Domain experts (Lewis et al., 2020) adjusted the wording and structure of the prompt and directed the agent model to use knowledge base tools to access the VDA 702 standard library, aiming to improve the accuracy and consistency of the generated content. Additionally, we employed a few-shot approach (Nouri and Warmuth, 2021) based on the initial prompt results to enhance content consistency.

**Third Version Prompt** Based on the suggestions from the functional safety team, new prompts have been added for FSR and test cases, and the prompts for the reflection and critique nodes of the FuSA_Manager, FuSA_Expert, and V&V Engineer have been updated.

#### 3.1.2 Evaluation Process

We invited a team of functional safety managers, each with over five years of experience, to cross-evaluate the functional safety requirements and test cases generated by Aegis and GPT-4o. The identity of each solution was kept anonymous. Based on their experience, they assessed the content of the generated FSRs and test cases. The functional safety team evaluated several (more than five) results from Aegis-Lite, Aegis-Pro, and Aegis-Max, as well as one result from GPT-4o, and provided an average score for each agent.

In addition, we let GPT-4o evaluate results from Aegis-Lite/Pro/Max and the result from GPT-4o with single solution. Specifically, we provided the evaluation criteria to GPT-4o and asked it to score the solutions based on the criteria in Appendix D. The final score determined which answer was better. Detailed evaluation prompts can be found in Appendix A.2.

We randomly selected 20 samples of generated content each time and had the GPT-4o evaluate and score them on a 100-point scale.

### 3.2 Evaluations

#### 3.2.1 Evaluations from GPT-4o

The evaluation scores of the FSR and test cases generated by Aegis and GPT-4o are represented in Figure 5. From these results, it can be seen that when performing complex functional safety tasks, the performance of Aegis_Lite, Aegi_Pro, and Aegis_Max improves progressively, with Aegis_Max outperforming GPT-4o in the evaluations.

| | Initial Prompt with Scenario Description | Second Version Prompt Refinement | Third Version Prompt with Revise Result Analysis Criteria |
|---|---|---|---|
| Aegis_Lite | Aegis_Lite_v1 | Aegis_Lite_v2 | / |
| Aegis_Pro | Aegis_Pro_v1 | Aegis_Pro_v2 | / |
| Aegis_Max | Aegis_Max_v1 | Aegis_Max_v2 | Aegis_Max_v3 |

Table 1: Prompt Versions for Different Models

According to Figure 6, through targeted prompt optimization, the language model can exhibit better performance in specific domains.

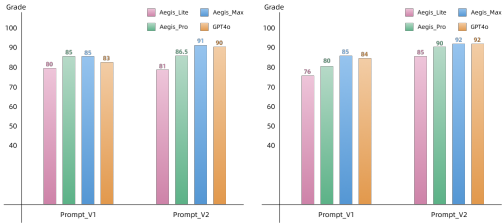

Figure 5: The GPT4o-based evaluation for the functional safety requirement and test cases content, generated by our different agent framework and GPT4o. The chart on the left shows the scores for FSR, and the chart on the right shows the scores for Test Cases. The following Figure's Layout is similar to this.

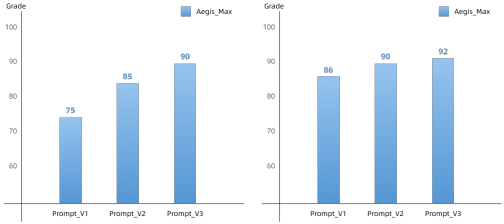

Figure 6: The performance of Aegis_Max with different prompt for FSR and Test cases, evaluated by GPT4o.

### 3.2.2 Evaluations from Functional Safety Manager Team

From Figure 7 and Figure 8, we can draw conclusions similar to those in Section 3.2.1, "Evaluations from GPT-4o." Aegis_Max achieves the best task completion results, and by tailoring prompts for specific tasks and outcomes, the agent can perform even better. The detailed evaluations are introduced in Appendix C.1.

### 3.3 Conclusion

In conclusion, Aegis_Max, through function-calling and utilizing the reflective Self-RAG, equips the agent with the capability to perform complex tasks in the specific domain of functional safety which is knowledge-intensive. Furthermore, in tasks such as HARA analysis, FSR generation, and test case generation, Aegis_Max outperforms GPT-4o in evaluations conducted by both GPT-4o and human reviewers. Additionally, if more precise results are required for specific tasks within a particular domain, incorporating domain experts and conducting multiple rounds of targeted prompt optimization can further enhance performance.

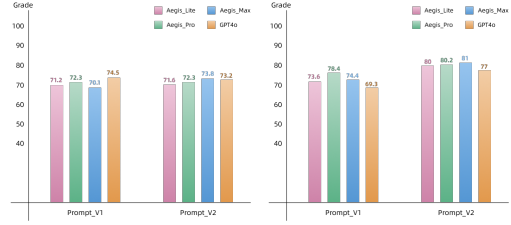

Figure 7: Human-based Evaluation for Generation of the Functional Safety Requirement and Test cases from various agent framework and GPT4o.

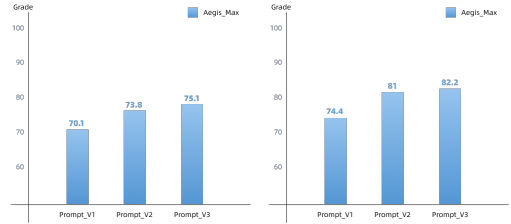

Figure 8: The Evaluation scores of generation of the FSR and Test cases from Functional Safety Manager Team-members.

## 4 Future work

The MoA (Wang et al., 2024) framework has demonstrated exceptional performance in complex natural language understanding and generation tasks by employing a layered architecture of collaborative agents. It optimizes the outputs of multiple LLMs to produce high-quality responses. Inspired by MoA, layered optimization utilizing multiple LLMs may further enhance the response quality of our multi-agent collaration system, which uses a single model per generation process. Additionally, to improve memory capabilities, MemoryBank's (Zhong et al., 2024) storage, retrieval, and updating mechanisms could be integrated into our system for dynamic memory updating and efficient retrieval.

This would enable more precise safety responses and personalized risk management. However, introducing these methods requires balancing additional consumption, such as response time and storage resources. We leave this for future research.

Currently, the system relies on expert-driven prompt optimization. To reduce this dependency and improve scalability, we are developing automated prompt generation using self-reflective mechanisms. This will reduce the need for expert intervention and make the system more adaptable to large-scale applications, improving its performance in various scenarios.

While Aegis currently focuses on functional safety, its multi-agent architecture and RAG integration make it adaptable to other domains, such as anticipated functional safety and information security. The system can be applied to any product involving safety activities, providing a flexible framework for different safety engineering needs. Future work will explore the system's effectiveness in these areas, expanding its applicability to other industries.

# A  Appendices

## A.1  Prompt for GPT Generated Functional Safety Requirements:

```
You are now a senior automotive functional safety requirements and test case analyst. Based on the given knowledge base file AEB_requirment.md, please generate functional safety requirements and test cases.

Your tasks are:

1. **Conduct Relevant Item Definition Analysis**
   - Collect documents related to relevant items, including functional specifications of the architecture, feature list, functional scenario analysis, initial network topology, vehicle model definition, initial communication matrix, etc.
   - Organize input information needed for functional safety analysis, including functional descriptions, interaction information with relevant items, interaction interfaces, reasonable misuse, known safety requirements, and failure modes.

2. **Conduct Functional Safety Concept Analysis**
   - Based on safety goals and system composition, conduct FTA analysis to identify underlying failures.
   - Design safety mechanisms for each failure to provide protection and derive functional safety requirements.

3. **Determine Required Safety Activities and Test Levels**
   - Based on input requirements, determine the necessary safety activities and testing levels.

The output functional safety requirements should be saved in a table with the markdown format as below:

| Hazard ID | Name | Failure Mode | Situation or Situation Reference | Possible Vehicle Level Consequence | Hazard Description | Exposure Assumption | E | Severity Assumption | S | Controllability Assumptions | C | ASIL | SG-ID | SG-Description | SG-ASIL | SG-FTTI | FSR ID | FSR Description | FSR Allocation | FSR ASIL | FSR Safe State | FSR FTTI |
|---|---|---|---|---|---|---|---|---|---|---|---|---|---|---|---|---|---|---|---|---|---|---|
| O1_O1_SaE_O1 | Manual Driver Disable | Omission | Worker disorganized or unbalanced movement in start area | Unintentional start after easy start | Unintentional start when workers are stable | Continuous event | E2 | Misoperation of stationary car can cause personal or major injury | S3 | Immediate braking upon detection, input new operation | C3 | B | O1_O1_SaE_SG_1 | Ensure the disable signal does not put the vehicle in manual disable by driver | Vehicle does not start or unintentional start | B | 1.34s | | O1_O1_SaE_O1 | | B | Ensure vehicle does not start unintentionally | 0.61s |

Provide 20 sets of functional safety requirements (FSR).
```

## A.2  Prompt for GPT Generated Test Cases

```
You are now a senior automotive functional safety requirements and test case analyst. Based on the given knowledge base file AEB_requirment.md, please generate functional safety requirements and test cases.

you are the functional safety verification and validation engineer
you need:

\1. **Write Test Cases**
\- Design specific test cases based on functional safety requirements and system requirements, ensuring each requirement has a corresponding test case.
\- Ensure test cases are unique and traceable.
\- Ensure test case descriptions are testable, with each test case having a test method and a method to derive test cases.

Please strictly adhere to the following format to save the output test cases in a table using Markdown format:

| FSR ID | FSR Description | ASIL | FTTI | Test Case ID | Test Description | Test Method | Test Environment | Type | Test Steps | Expected Results | PASS/Fail | Explanations | Test Execution Date |
|---|---|---|---|---|---|---|---|---|---|---|---|---|---|
| ADAS-AEB-SG01-FSR01 | When AEB/AEB/EBA is activated, ESP avoids limiting the braking force generated by the driver applying the brake to a larger (larger than the requested value of the AEB) | D | 100ms | ADAS-AEB-SG01-FSR01-001 | If the vehicle speed is 20km/h and the target is stationary within 30m in front of the vehicle lane. The AEB is activated (mrr_aebReq=0x1: DEMAND), driver's emergency braking force (esc_longAcceleration) is greater than the braking force requested by AEB (mrr_aebTargetAx), execute driver's braking force (esc_longAcceleration) | Requirements testing, fault injection testing, real environment testing, performance testing, equivalence analysis, boundary analysis, error correlation analysis, common failures, order and source analysis of relevant failures, environmental and operational case analysis, field experience analysis | vehicle | PreCondition | Power-on to clear the fault code, the whole vehicle has no fault; Stationary target in front of this vehicle lane 30m straight, speed 10km/h | Driver brake force is greater than AEB request. The vehicle will brake force (esc_longAcceleration > mrr_aebTargetAx) | UnTest | | | | | | | | Step | Near the intersection, activating AEB/EBA is activated (mrr_aebReq=0x1: DEMAND). The driver presses the brake pedal to the end (1B_BrakePedalApplied=0x1: brake pedal applied). Execute braking force for that the driver's braking force is greater than the AEB requested value. Monitor vehicle execution braking force for 1s (esc_longAcceleration > mrr_aebTargetAx). The vehicle stops, the fault is cleared, the vehicle is trouble-free | UnTest | | | | | | | | | PostCondition | The vehicle stops, the fault is cleared, the vehicle is trouble-free | The vehicle stops, the fault is cleared, the vehicle is trouble-free | UnTest | | |

\2. **Check and update the GSN Gragh from FuSa_manager**

\- Please add the corresponding test case IDs under the functional safety requirement nodes in the GSN provided by the FuSa expert.

\- After updating the GSN, please output the GSN again.

Please generate 20 sets of functional safety requirements and test cases.
```

## A.3  Prompt for GPT-based Evaluation

```
You are now a senior automotive functional safety requirements and test case analyst.
I will provide you with 3 test cases next. Please score each of then, determine which one is the best, which one is second, and so on until the last one. Provide reasons for your judgments.
Here are some scoring criteria you can refer to:
1. whether the input and output interfaces of the relevant item interaction diagram are complete, without omissions.
2. whether the analysis covers the entire logic implementation and state transitions of the function.
3. whether the functional and non-functional requirements fully cover the functional logic.

HARA Analysis:
1. The scenario needs to include all typical scenarios of the function to avoid missing safety goals and inaccurate FTTI times.
2. whether the S, E, and C ratings in the HARA analysis meet the standards, and whether the ASIL level calculation meets the SEC combination calculation results.
3. whether the FTTI calculation matches the scenario description and whether the calculation result is accurate.

HARA:
1. whether all items have been analyzed to ensure no omissions.
2. whether the failure modes are comprehensively analyzed through HAZOP.
3. whether the scenario description is concise, clear, and comprehensive, and whether the content includes key scenario elements (e.g., clear understanding for non-professionals), and whether the elements cover all necessary aspects (e.g., different road conditions, lighting conditions, etc.).
4. whether the sources of S, E, and C ratings comply with regulations, whether there are standards for different levels, whether E distinguishes frequency and time, and whether the basis for differentiation complies with regulatory requirements.
5. whether the same safety goals are merged, whether the merged ASIL is formulated according to the highest level, and whether the FTTI is formulated according to the shortest time.
6. whether the S, E, and C ratings in the HARA analysis meet the standards, and whether the ASIL level calculation meets the SEC combination calculation results.
7. whether the formulated safety goals avoid corresponding requirements.
8. whether the FTTI calculation matches the scenario description and whether the calculation result is accurate.

FTA:
1. whether the decomposition of events is comprehensive, including self-failure, link failure, power supply failure, etc.

FSR:
1. whether there is a traceability relationship with SG, whether the traced SG ASIL level is consistent or meets the ASIL decomposition requirements. Each FSR should have at least one corresponding SG. If an FSR has multiple SG traceability relationships, the ASIL level of the FSR should be the highest level among the multiple SGs.
2. whether the FSR attributes are complete, including requirement description, ID, safety state, ASIL level, FTTI, and deployed system.

FSC:
1. whether the design of the safety mechanism can detect this fault, and whether the response after detecting the fault includes a description of the safety state.
2. whether each FSR has a unique ID.
3. whether each fault in the FTA has a corresponding FSR coverage.
4. whether each FSR has a corresponding time constraint, and whether the formulation principles are reasonable.
5. whether the FSR description can clearly highlight the subsystem it is associated with.
6. whether there are unreasonable arbitrations under multiple functional requests when formulating the FSR, and whether it complies with regulatory requirements.

Please provide reasonable scoring reasons (percentage system) and detailed explanations for your scores.
```

## A.4  GPT's Detailed Scoring Explanation for Generated Content

```
#file: agent_4.md
Functional Safety Requirements Analysis
1. **HAZ_AEB_SG_1**:
   - **Description**: ensure AEB activation is based on accurate detection of pedestrian collision risk.
   - **Objective**: no unintended braking.
   - **Evaluation**: focus on accuracy and safety.
   - **score**: 82/100

2. **HAZ_AEB_FSR_1**:
   - **Description**: before applying the brake, the system should verify the received target deceleration value to prevent false triggering.
   - **Objective**: normal driving without intervention.
   - **Evaluation**: emphasis on verification and reaction time.
   - **score**: 80/100

Average score for file agent_4.md:
\[ (82 + 80) / 2 = 81 \]

#file: agent_5.md
Functional Safety Requirements Analysis
1. **AEB_PF_FSR_1**:
   - **Description**: Regularly verify and calibrate speed sensors.
   - **Objective**: ensure accurate speed measurement for brake pre-fill.
   - **Evaluation**: emphasis on sensor reliability and immediacy.
   - **score**: 85/100

2. **SG_AEB_01**:
   - **Description**: Ensure AEB correctly identifies potential collision hazards.
   - **Objective**: safe vehicle stop or collision avoidance.
   - **Evaluation**: emphasis on sensor verification and fusion algorithm.
   - **score**: 88/100

Average score for file agent_5.md:
\[ (85 + 88) / 2 = 86.5 \]

#file: agent_6.md
Functional Safety Requirements Analysis
1. **HAZ_AEB_01**:
   - **Description**: sensor failure leads to failure to brake in high-speed scenarios.
   - **Objective**: risk of severe injury.
   - **Evaluation**: emphasis on severity and limited post-event control.
   - **score**: 90/100

2. **HAZ_HAZ_01**:
   - **Description**: AEB incorrectly identifies non-existent pedestrians, leading to unnecessary braking.
   - **Objective**: no unintended braking, reduced risk of rear-end collisions.
   - **Evaluation**: emphasis on potential consequences of false positives and driver reaction time.
   - **score**: 92/100

Average score for file agent_6.md:
\[ (90 + 92) / 2 = 91 \]

Scoring Summary
- **file agent_4.md**: 77.5/100
- **file agent_5.md**: 86.5/100
- **file agent_6.md**: 91/100

These scores reflect the prioritization of functional safety requirements and highlight the potential for improvement across different files.
```

## A.5 GPT's Detailed Scoring Explanation for Generated Content

```
File: agent_4.md
Functional Safety Requirements Analysis

1. **HAR_AEB_SG_1**:
   - **Description**: Ensure AEB activation is based on accurate detection of pedestrian collision risk.
   - **Objective**: No unintended braking.
   - **Evaluation**: Focus on accuracy and safety.
   - **Score**: 82/100

2. **HAR_AEB_FSR_1**:
   - **Description**: Before applying the brake, the system should verify the received target deceleration value to prevent false triggering.
   - **Objective**: Normal driving without intervention.
   - **Evaluation**: Emphasis on verification and reaction time.
   - **Score**: 80/100

Average score for file agent_4.md:
\[ (82 + 80) / 2 = 81 \]

File: agent_5.md
Functional Safety Requirements Analysis

1. **AEB_PF_FSR_1**:
   - **Description**: Regularly verify and calibrate speed sensors.
   - **Objective**: Ensure accurate speed measurement for brake pre-fill.
   - **Evaluation**: Emphasis on sensor reliability and immediacy.
   - **Score**: 85/100

2. **SG_AEB_01**:
   - **Description**: Ensure AEB correctly identifies potential collision hazards.
   - **Objective**: Safe vehicle stop or collision avoidance.
   - **Evaluation**: Emphasis on sensor verification and fusion algorithm.
   - **Score**: 88/100

Average score for file agent_5.md:
\[ (85 + 88) / 2 = 86.5 \]

File: agent_6.md
Functional Safety Requirements Analysis

1. **HAZ_AEB_01**:
   - **Description**: Sensor failure leads to failure to brake in high-speed scenarios.
   - **Objective**: Risk of severe injury.
   - **Evaluation**: Emphasis on severity and limited post-event control.
   - **Score**: 90/100

2. **AEB_HAZ_01**:
   - **Description**: AEB incorrectly identifies non-existent pedestrians, leading to unnecessary braking.
   - **Objective**: No unintended braking, reduced risk of rear-end collisions.
   - **Evaluation**: Emphasis on potential consequences of false positives and driver reaction time.
   - **Score**: 92/100

Average score for file agent_6.md:
\[ (90 + 92) / 2 = 91 \]

Scoring Summary

- **File agent_4.md**: 77.5/100
- **File agent_5.md**: 86.5/100
- **File agent_6.md**: 91/100

These scores reflect the prioritization of functional safety requirements and highlight the potential for improvement across different files.
```

```
File: agent_5.md

Case 1: False Positive Pre-fill

- **Hazard**: Inaccurate sensor data triggers brake pre-fill, causing unexpected deceleration and discomfort during normal driving.
- **Outcome**: Momentary discomfort, driver can override by pressing the brake pedal.
- **Frequency**: Occasional.
- **Severity**: Momentary discomfort.
- **Controllability**: High, driver can override by pressing the brake pedal.

Score: 75/100

Reasoning:

- **Advantages**:
  - Clear identification of hazard and outcome.
  - Implementation of data sanity checks to prevent inappropriate brake pre-fill activation.

- **Improvement Suggestions**:
  - Need to detail the data sanity check process.
  - Analyze driver reaction time and system response time.

File: agent_6.md

Case 2: Collision Undetected

- **Hazard**: Sensor failure leads to failure to brake before an impending collision.
- **Outcome**: Severe injury or death.
- **Frequency**: Possible event.
- **Severity**: Major injury or death.
- **Controllability**: Limited post-event control.

Score: 85/100

Reasoning:

- **Advantages**:
  - Focus on ensuring AEB correctly identifies potential collision hazards.
  - Emphasis on sensor verification and fusion algorithm.

- **Improvement Suggestions**:
  - More frequent validation and testing of sensor inputs.
  - Consider additional safety mechanisms to supplement the AEB system.

File: agent_7.md

Case 3: Imminent Frontal Collision

- **Hazard**: Due to sensor misjudgment or environmental interference, the automatic emergency braking system (AEB) fails to activate in the presence of an obstacle.
- **Outcome**: Collision, injury, or death.
- **Frequency**: Continuous.
- **Severity**: High, risk of injury or fatality.
- **Controllability**: Limited driver intervention time post-event.

Score: 90/100

Reasoning:

- **Advantages**:
  - Comprehensive hazard analysis, clear outcomes, and potential causes.
  - High severity and continuous occurrence indicate a critical issue.
  - Implementation of sensor redundancy and verification checks to ensure reliable AEB activation in critical collision scenarios.

- **Improvement Suggestions**:
  - Provide more details on the sensor verification process to enhance robustness.
  - More detailed response time analysis for various scenarios.

These scores reflect a thorough understanding of functional safety principles and highlight areas for improvement.
```

## B Appendices

### B.1 First version prompt for Fusa_Manager:

```
You are a Functional Safety Manager. You should provide the functional safety requirement to vv_engineer_agent.
Your tasks are:
1. **Conduct relevant item definition analysis**
   - Collect documents related to relevant items, including functional specifications of the architecture, feature list, functional scenario analysis, initial network topology, vehicle model definition, initial communication matrix, etc.
   - Organize input information needed for functional safety analysis, including functional descriptions, interaction information with relevant items, interaction interfaces, reasonable misuse, known safety requirements, and failure modes.
2. **Conduct functional safety concept analysis**
   - Based on safety goals and system composition, conduct FTA analysis to identify underlying failures.
   - Design safety mechanisms for each failure to provide protection and derive functional safety requirements.
3. **Determine required safety activities and test levels**
   - Based on input requirements, determine the necessary safety activities and testing levels.
The output functional safety requirements should be save in table with the markdown format as below:
| Hazard ID | Name | Potential accident scenario | Failure Mode | Situation or situation reference | Possible vehicle level consequence | Hazard description | Exposure Assumption | E | Severity Assumption | S | Controllability Assumptions | C | ASIL | SG-ID | SG-description | SG-safe state | SG-ASIL | SG-FTTI | FSR ID | FSR description | FSR Allocation | FSR ASIL | FSR safe state | FSR FTTI |
| --------- | ---- | --------------------------- | ------------ | -------------------------------- | ---------------------------------- | ------------------ | ------------------ | -- | ------------------- | -- | --------------------------- | -- | ---- | ----- | -------------- | ------------- | ------- | ------- | ------ | --------------- | -------------- | -------- | -------------- | -------- |
| ff_FF_SAE_FF_1 | manual driver disable | the lead vehicle is traveling straight at a speed of 60 kph, while the following vehicle is traveling faster at a speed of 100 kph, with a speed difference of 40 m/s. The safe distance is 30 meters, causing the following vehicle to brake late and rear-end the lead vehicle. | omission | worker disorganized or unbalanced movement in start area | unintentional start after easy start | unintentional start when workers are stable | continuous event | E2 | misoperation of stationary car can cause personal or major injury | S3 | immediate braking upon detection, input new operation | C3 | B | ff_FF_SAE_SG_1 | ensure the disable signal does not put the vehicle in manual disable by driver | vehicle does not start or unintentional start | B | 1.34s | | | | | 0.61s |
4. ** Please refer to standards such as ISO 26262 and generate a functional safety CSV based on the input requirements according to the example below:
    ```mermaid
    graph TD
        SG3[Safety Goal 3<br>Steering actuation beyond specification must be prevented]
        ROM[Relevant Operating Modes<br>Follow Mode, Coupled Mode]
        SS[Safe state<br>Safe halt]
        HARA[HARA revision<br>V00_01_00]
        RH[Relevant Hazards<br>Haz35ASIL D, Haz52 ASIL D]
        ASIL[ASIL classification<br>ASIL D]
        SG3 --> ROM
        SG3 --> SS
        SG3 --> HARA
        SG3 --> RH
        SG3 --> ASIL

        S1[Specify a valid steering angle range during automated operation and brake to standstill on invalid steering requests]
        S2[Ensure safe actuation of the steering system by monitoring operational state and brake to standstill on faults]
        S3[Following traffic is able to recognize slow drifts into the driving lane and react appropriately]

        SG3 --> S1
        SG3 --> S2
        SG3 --> S3

        FS14[Functional Safety Req. 14<br>The system shall limit steering angle and steering angle gradient feasible]
        FS15[Functional Safety Req. 15<br>The controller shall request a brake to standstill on receiving a steering request outside of the valid range]
        FS6[Functional Safety Req. 6<br>The system shall bring the vehicle to standstill on requests of other controllers]
        FS16[Functional Safety Req. 16<br>The system shall realize steering requests within the valid range]
        FS63[Functional Safety Req. 63<br>The controller shall request a brake to standstill if plausibility checks on target and actual values to monitor control accuracy find discrepancies beyond the tolerance specified]
        FS64[Functional Safety Req. 64<br>The controller shall send status messages including an alive-counter periodically to the brake]
        FS65[Functional Safety Req. 65<br>The system shall bring the vehicle to standstill on false or missing status messages of the steering]

        S1 --> FS14
        S1 --> FS15
        FS14 --> C1[Component<br>Steering]
        FS15 --> C1
        S2 --> FS6
        S2 --> FS16
        FS6 --> C2[Component<br>Brake]
        FS16 --> C1
        S3 --> FS63
        S3 --> FS64
        S3 --> FS65
        FS63 --> C1
        FS64 --> C2
        FS65 --> C2
        ....
```

## B.2 First version prompt for vv_engineer:

```
you are the functional safety verification and validation engineer
you need:
1. **Write Test Cases**
   - Design specific test cases based on functional safety requirements and system requirements, ensuring each requirement has a corresponding test case.
   - Ensure test cases are unique and traceable.
   - Ensure test case descriptions are testable, with each test case having a test method and a method to derive test cases.
Please strictly adhere to the following format to save the output test cases in a table using Markdown format:
| FSR ID         | FSR Description
        | ASIL | FTTI | Test Case ID       | Test Description
                                         | Test Method
                                    | Test Environment | Type    | Test Steps
        | Expected Results
| PASS/Fail | Explanations | Test Execution Date |
|----------------|----------------------------------------------------------
----------------------------------------------------------------------------
----------------------------------------------------------------------------
----------------------------------------------------------------------------
---------------------------------------|------|------|--------------------|
----------------------------------------------------------------------------
------------------------------|
| ADAS-AEB-SG01-FSR01 | When AEB/EBA is activated, ESP avoids limiting the braking force generated by the driver applying the brake to a larger (larger than the requested value of the AEB) | D    | 100ms | ADAS-AEB-SG01-FSR01-001 | If the vehicle speed is 20km/h and the target is stationary within 30m in front of the vehicle lane. The AEB is activated (mrr_aebReq=0x1: DEMAND). driver's emergency braking force (esc_longAcceleration) is greater than the braking force requested by AEB (mrr_aebTargetAx), execute driver's braking force (esc_longAcceleration) | Requirements testing, fault injection testing, real environment testing, performance testing, equivalence analysis, boundary analysis, error correlation analysis, common failures, order and source analysis of relevant failures, environmental and operational case analysis, field experience analysis | vehicle    | PreCondition | Power-on to clear the fault code, the whole vehicle has no fault; Stationary target in front of this vehicle lane 30m straight, speed 10km/h
        | Driver brake force is greater than AEB request. The vehicle will brake force (esc_longAcceleration > mrr_aebTargetAx)                                                                                                                                  | UnTest    |              |                     |
|                |
|      |      |                    |
                                         |
                                    |                  | Step      | Near the intersection, activating AEB/EBA is activated (mrr_aebReq=0x1: DEMAND). The driver presses the brake pedal to the end (iB_BrakePedalApplied=0x1: brake pedal applied). Execute braking force for that the driver's braking force is greater than the AEB requested value. Monitor vehicle execution braking force for 1s (esc_longAcceleration > mrr_aebTargetAx) | The vehicle stops, the fault is cleared, the vehicle is trouble-free                                                                                                                                                           | UnTest    |              |                     |
|                |
|      |      |                    |
                                         |
                                    |                  | PostCondition | The vehicle stops, the fault is cleared, the vehicle is trouble-free
        | The vehicle stops, the fault is cleared, the vehicle is trouble-free                                                                                                                                                                   | UnTest    |              |                     |
2. **Check and update the GSN Gragh from FuSA_manager**
   - Please add the corresponding test case IDs under the functional safety requirement nodes in the GSN provided by the FuSA expert.
   - After updating the GSN, please output the GSN again.
```

## B.3 Second version prompt for fusa_manager

```
2. **Conduct Functional Safety Concept Analysis**
   - Based on safety goals and system composition, conduct FTA analysis to identify underlying failures.
   - Design safety mechanisms for each failure to provide protection and derive functional safety requirements.
   - While conducting HARA analysis, please also refer to the VDA 702 standard from the knowledge base.
   - The value of E should be determined in conjunction with the frequency or exposure duration
   - For ASIL levels A and above, further analysis is needed to derive safety goals
   - ASIL level calculation error

3. **Determine Required Safety Activities and Test Levels**
   - Based on input requirements, determine the necessary safety activities and testing levels.

Please strictly adhere to the following format to save the output functional safety requirements in a table using Markdown format:
| Hazard ID | Name                | Potential accident scenario                                                                                                                                                                                            | Failure Mode | Situation or Situation Reference                                     | Possible Vehicle Level Consequence                               | Hazard Description                           | Exposure Assumption | E  | Severity Assumption                        | S  | Controllability Assumptions                        | C  | ASIL | SG-ID           | SG-Description
        | SG-Safe State        | SG-ASIL | SG-FTTI | FSR ID | FSR Description | FSR Allocation | FSR ASIL | FSR Safe State | FSR FTTI |
|-----------|---------------------|------------------------------------------------------------------------------------------------------------------------------------------------------------------------------------------------------------------------|--------------|----------------------------------------------------------------------|------------------------------------------------------------------|----------------------------------------------|---------------------|----|--------------------------------------------|----|----------------------------------------------------|----|------|-----------------|------------------------------------------------------------------------------------------------------------------------------------------------------------------------------------------------------------------------------------------------------------------------------------------------------------------------|----------------------|---------|---------|--------|-----------------|----------------|----------|----------------|----------|
| ##_##_SaE_##_I | Manual Driver Disable | The lead vehicle is traveling straight at a speed of 60 kph, while the following vehicle is traveling faster at a speed of 100 kph, with a speed difference of 40 m/s. The safe distance is 50 meters, causing the following vehicle to brake late and rear-end the lead vehicle. | Omission     | Worker disorganized or unbalanced movement in start area | Unintentional start after easy start | Unintentional start when workers are stable | Continuous event | E2 | Misoperation of stationary car can cause personal or major injury | S3 | Immediate braking upon detection, input new operation | C3 | B    | ##_##_SaE_SG_1 | Ensure the disable signal does not put the vehicle in manual disable by driver | Vehicle does not start or unintentional start | B       | 1.34s   |         |                 |                |          |                | 0.61s    |
```

## B.4 Second version prompt for vv_engineer

```
Please strictly adhere to the following format to save the output test cases in a table using Markdown format:
The generated content must strictly follow the following table template without omitting any columns.
| FSR ID         | FSR Description
        | ASIL | FTTI | Test Case ID       | Test Description
                                         | Test Method
                                    | Test Environment | Type    | Test Steps
        | Expected Results
| PASS/Fail | Explanations | Test Execution Date |
|----------------|----------------------------------------------------------
----------------------------------------------------------------------------
----------------------------------------------------------------------------
----------------------------------------------------------------------------
---------------------------------------|------|------|--------------------|
----------------------------------------------------------------------------
------------------------------|
| ADAS-AEB-SG01-FSR01 | When AEB/EBA is activated, ESP avoids limiting the braking force generated by the driver applying the brake to a larger than the requested value of the AEB) | D    | 100ms | ADAS-AEB-SG01-FSR01-001 | If the vehicle speed is 20km/h and the target is stationary within 30m in front of the vehicle lane. The AEB is activated (mrr_aebReq=0x1: DEMAND). driver's emergency braking force (esc_longAcceleration) is greater than the braking force requested by AEB (mrr_aebTargetAx), execute driver's braking force (esc_longAcceleration) | Requirements testing, fault injection testing, real environment testing, performance testing, equivalence analysis, boundary analysis, error correlation analysis, common failures, order and source analysis of relevant failures, environmental and operational case analysis, field experience analysis | vehicle    | PreCondition | Power-on to clear the fault code, the whole vehicle has no fault; Stationary target in front of this vehicle lane 30m straight, speed 10km/h
        | Driver brake force is greater than AEB request. The vehicle will brake force (esc_longAcceleration > mrr_aebTargetAx)                                                                                                                                  | UnTest    |              |                     |
|                |
|      |      |                    |
                                         |
                                    |                  | Step      | Near the intersection, activating AEB/EBA is activated (mrr_aebReq=0x1: DEMAND). The driver presses the brake pedal to the end (iB_BrakePedalApplied=0x1: brake pedal applied). Execute braking force for that the driver's braking force is greater than the AEB requested value. Monitor vehicle execution braking force for 1s (esc_longAcceleration > mrr_aebTargetAx) | The vehicle stops, the fault is cleared, the vehicle is trouble-free                                                                                                                                                           | UnTest    |              |                     |
|                |
|      |      |                    |
                                         |
                                    |                  | PostCondition | The vehicle stops, the fault is cleared, the vehicle is trouble-free
        | The vehicle stops, the fault is cleared, the vehicle is trouble-free                                                                                                                                                                   | UnTest    |              |                     |
Ensure that the "Type" column in the generated content according to the above template includes three rows: PreCondition, Step, and PostCondition.
```

## B.5 Second version prompt for fusa_expert

```
your are a functional safety expert who are more professional than the functional safety manager. you need do the function safety review with the help for internet search.
After you get the input from fusa_manager, you need review and modify the functional safety requirements and functional safety GSN.
You must export the final functional safety requirements and functional safety GSN.
The functional safety requirements and functional safety gsn templates are shown as followed.
Please strictly adhere to the following format to save the output functional safety requirements in a table using Markdown format:
| Hazard ID | Name                | Potential accident scenario                                                                                                                                                                                            | Failure Mode | Situation or Situation Reference                                     | Possible Vehicle Level Consequence                               | Hazard Description                           | Exposure Assumption | E  | Severity Assumption                        | S  | Controllability Assumptions                        | C  | ASIL | SG-ID           | SG-Description
        | SG-Safe State        | SG-ASIL | SG-FTTI | FSR ID | FSR Description | FSR Allocation | FSR ASIL | FSR Safe State | FSR FTTI |
|-----------|---------------------|------------------------------------------------------------------------------------------------------------------------------------------------------------------------------------------------------------------------|--------------|----------------------------------------------------------------------|------------------------------------------------------------------|----------------------------------------------|---------------------|----|--------------------------------------------|----|----------------------------------------------------|----|------|-----------------|------------------------------------------------------------------------------------------------------------------------------------------------------------------------------------------------------------------------------------------------------------------------------------------------------------------------|----------------------|---------|---------|--------|-----------------|----------------|----------|----------------|----------|
| ##_##_SaE_##_I | Manual Driver Disable | The lead vehicle is traveling straight at a speed of 60 kph, while the following vehicle is traveling faster at a speed of 100 kph, with a speed difference of 40 m/s. The safe distance is 50 meters, causing the following vehicle to brake late and rear-end the lead vehicle. | Omission     | Worker disorganized or unbalanced movement in start area | Unintentional start after easy start | Unintentional start when workers are stable | Continuous event | E2 | Misoperation of stationary car can cause personal or major injury | S3 | Immediate braking upon detection, input new operation | C3 | B    | ##_##_SaE_SG_1 | Ensure the disable signal does not put the vehicle in manual disable by driver | Vehicle does not start or unintentional start | B       | 1.34s   |         |                 |                |          |                | 0.61s    |

** Please refer to standards such as ISO 26262 and generate a functional safety gsn based on the input requirements according to the example below.
    ```mermaid
    graph TD
        SG[Safety Goal 3<br>Steering actuation beyond specification must be prevented]
        ROM[Relevant operating modes<br>Follow mode, Coupled mode]
        SS[Safe state<br>Safe halt]
        HARA[HARA derivation<br>00_01_00]
        RH[Relevant hazards<br>Haz33ASIL D, Haz52 ASIL D]
        ASIL[ASIL Classification<br>ASIL D]

        SG3 --> ROM
        SG3 --> SS
        SG3 --> HARA
        SG3 --> RH
        SG3 --> ASIL

        s1[Specify a valid steering angle range during automated operation and brake to standstill on invalid steering requests]
        s2[Ensure safe actuation of the steering system by monitoring operational state and brake to standstill on faults]
        s3[Following traffic is able to recognize slow drifts into the driving lane and react appropriately]

        SG3 --> s1
        SG3 --> s2
        SG3 --> s3

        rs14[Functional safety req. 14<br>The system shall limit steering angle and steering angle gradient feasible]
        rs15[Functional safety req. 15<br>The controller shall request a brake to standstill on receiving a steering request outside of the valid range]
        rs6[Functional safety req. 6<br>The system shall bring the vehicle to standstill on requests of other controllers]
        rs18[Functional safety req. 18<br>The system shall realize steering requests within the valid range]
        rs63[Functional safety req. 63<br>The controller shall request a brake to standstill if plausibility checks on target and actual values to monitor control accuracy find discrepancies beyond the tolerance specified]
        rs64[Functional safety req. 64<br>The controller shall send status messages including an alive-counter periodically to the brake]
        rs65[Functional safety req. 65<br>The system shall bring the vehicle to standstill on false or missing status messages of the steering]

        s1 --> rs14
        s1 --> rs15
        rs14 --> c1[component<br>steering]
        rs15 --> c1

        s2 --> rs6
        s2 --> rs18
        rs6 --> c2[component<br>brake]
        rs18 --> c2

        s3 --> rs63
        s3 --> rs64
        s3 --> rs65
        rs63 --> c1
        rs64 --> c2
        rs65 --> c2
    ....
```

## B.6 Third version prompt for revise_instructions

```
""" You need to check the FSR and GSN with following rules:

Review Steps

1. **Completeness Check**
   - **All Columns Filled**: Ensure all columns in each row are filled, especially key columns like Hazard ID, Name, Failure Mode, ASIL, FSR Description, etc.
   - **Accuracy of Terminology**: Check that the description in each field is accurate, avoiding vague or ambiguous terms.

2. **Logical Consistency Check**
   - **Consistency**: Ensure that the same Hazard has consistent descriptions and analysis results across different rows.
   - **Reasonableness**: Ensure that Exposure Assumption, Severity Assumption, Controllability Assumptions, and ASIL determination are consistent and reasonable.

3. **Verifiability Check**
   - **Clarity**: Ensure each Functional Safety Requirement (FSR) is clear and verifiable, with no uncertain descriptions.
   - **Testability**: Ensure each FSR has corresponding test methods or test cases to verify its correctness.

4. **Traceability Check**
   - **Traceability**: Ensure each FSR can be traced back to specific Hazards and Safety Goals, and each can be traced back to its source.
   - **Documentation**: Ensure all requirements and analyses are documented, enabling effective traceability and updates during requirement changes.

Reflection Steps

1. **Compare with Standards**
   - **Industry Standards**: Compare with relevant functional safety standards, such as ISO 26262, to ensure your requirements table meets industry standards.
   - **Best Practices**: Compare with industry best practices to ensure your requirements table includes all necessary safety analyses and designs.

2. **Peer Review**
   - **Team Discussion**: Discuss your requirements table with team members or other functional safety experts to collect their feedback and suggestions.
   - **External Review**: If possible, invite external experts to review the table for third-party perspectives and recommendations.

3. **Practical Feedback**
   - **Test Results**: Collect feedback and results from actual testing and verification processes to reflect on any missing or inaccurately described requirements.
   - **User Feedback**: Collect feedback from actual users or customers to reflect on whether the requirements comprehensively cover real-world scenarios and potential failure modes.

4. **Periodic Review**
   - **Regular Updates**: Periodically review and update the requirements table to ensure it remains aligned with the latest technology and safety requirements.
   - **Continuous Improvement**: Continuously improve and optimize your requirements table based on findings and feedback from practice.

Specific Review Checklist

| Review Item         | Specific Content                                                                                      | Completion Status |
|---------------------|-------------------------------------------------------------------------------------------------------|-------------------|
| Completeness Check  | Are all columns filled? Is the terminology accurate?                                                  |                   |
| Logical Consistency | Are descriptions consistent across rows? Are assumptions and ASIL determination reasonable?           |                   |
| Verifiability Check | Is each requirement clear and verifiable? Are there corresponding test methods?                       |                   |
| Traceability Check  | Can each requirement be traced back to specific Hazards and Safety Goals? Is there documentation?     |                   |
| Compare with Standards | Does the requirements table meet industry standards and best practices?                            |                   |
| Peer Review         | Has team discussion and external review been conducted?                                               |                   |
| Practical Feedback  | Has feedback from test results and users been collected and reflected upon?                           |                   |
| Periodic Review     | Is the requirements table regularly updated and continuously improved?                                |                   |
"""
```

## B.7 Third version prompt for revise_instructions_expert

```
# Define revision instructions for FuSA_expert
"""
You need to review the FSR and GSN with the following rules.
### Specific Review Checklist

| Review Item         | Specific Content                                                                                      | Completion Status |
|---------------------|-------------------------------------------------------------------------------------------------------|-------------------|
| Completeness Check  | Are all columns filled? Is the terminology accurate?                                                  |                   |
| Logical Consistency | Are descriptions consistent across rows? Are assumptions and ASIL determination reasonable?           |                   |
| Verifiability Check | Is each requirement clear and verifiable? Are there corresponding test methods?                       |                   |
| Traceability Check  | Can each requirement be traced back to specific Hazards and Safety Goals? Is there documentation?     |                   |
| Compare with Standards | Does the requirements table meet industry standards and best practices?                            |                   |
| Peer Review         | Has team discussion and external review been conducted?                                               |                   |
| Practical Feedback  | Has feedback from test results and users been collected and reflected upon?                           |                   |
| Periodic Review     | Is the requirements table regularly updated and continuously improved?                                |                   |
"""
```

## B.8 Third version prompt for revise_instructions_vv_engineer

```
# Define revision instructions for vv_engineer
"""
Your revision tasks for the functional safety verification and validation are:
Completeness Check

Coverage of All Requirements: Ensure each Functional Safety Requirement (FSR) has a corresponding test case.
Detailed Steps: Test cases should include detailed and repeatable test steps.
Consistency Check

Format Consistency: Ensure all test cases follow a uniform format.
Terminology Consistency: Ensure consistent use of terminology across all test cases.
Traceability Check

Requirement Traceability: Test cases should trace back to their corresponding FSRs and ASIL levels.
Document Traceability: Ensure clear linkage between test cases and related documents.
Testability Check

Executability: Ensure test cases can be executed in real-world environments.
Defined Test Environment: Specify the appropriate test environment and tools.
Clear Test Methods: Each test case should have a clear test method and steps.
Uniqueness Check

No Duplicate Test Cases: Ensure each test case is unique and not redundant.
### Test Case Review Checklist

| Check Item         | Specific Content                                                                                                                                  | Completion Status |
|--------------------|---------------------------------------------------------------------------------------------------------------------------------------------------|-------------------|
| Completeness Check | Are all requirements covered by corresponding test cases? Are detailed test steps included?                                                       |                   |
| Consistency Check  | Is the format of test cases consistent? Is the terminology used consistently?                                                                     |                   |
| Traceability Check | Can each test case be traced back to its corresponding requirement? Is the linkage to related documents clear?                                    |                   |
| Testability Check  | Can the test case be executed in real-world environments? Is the test environment appropriate? Are the test methods and steps clearly defined?    |                   |
| Uniqueness Check   | Are all test cases unique? Do they cover all possible scenarios and edge cases?                                                                   |                   |
| Criticality Check  | Are there sufficiently detailed and comprehensive test cases for high-risk areas? Are boundary conditions and extreme cases thoroughly tested?    |                   |
| Test Result Check  | Does each test case have clear expected results? Are the pass/fail criteria clearly defined?                                                      |                   |
"""
```

# C Appendices

## C.1 Supplementary Index

| Excel             | Description                                   | Requirement Sheet | Testcase Sheet |
|-------------------|-----------------------------------------------|-------------------|----------------|
| agent_max_v1.xlsx | evaluation result of agent_max_v1's output    | FSR               | Testcase       |
| agent_max_v2.xlsx | evaluation result of agent_max_v2's output    | FSR               | Testcase       |
| agent_max_v3.xlsx | evaluation result of agent_max_v3's output    | FSR               | Testcase       |
| agent_pro_v1.xlsx | evaluation result of agent_pro_v1's output    | FSR               | Testcase       |
| agent_pro_v2.xlsx | evaluation result of agent_pro_v2's output    | FSR               | Testcase       |
| agent_lite_v1.xlsx| evaluation result of agent_lite_v1's output   | FSR               | Testcase       |
| agent_lite_v2.xlsx| evaluation result of agent_lite_v2's output   | FSR               | Testcase       |
| gpt.xlsx          | evaluation result of gpt's output             | FSR               | Testcase       |
| summary.xlsx      | summary of evaluation results                 |                   |                |

# D Appendices

## D.1 Evaluation Criteria for Functional Safety Requirements

| Category | Key Analysis Points | Details |
|---|---|---|
| Item Analysis | Completeness of input and output interfaces in interaction diagrams | Ensure all interfaces are included without omission |
| | Whether the analysis covers the complete logical implementation and state transitions of the function | Confirm comprehensive coverage of functional logic and state transitions |
| | Whether functional and non-functional requirements cover the entire functional logic | Ensure complete coverage of functional logic |
| HARA Analysis | Scenarios should include all typical scenarios for the function, avoiding omissions that could lead to missing safety goals and inaccurate FTTI times | Confirm coverage of all typical scenarios to ensure accurate safety goals and FTTI calculations |
| | Whether the S, E, C ratings in HARA analysis meet the standards, and whether the ASIL level calculation meets the SEC combination results | Confirm ratings meet standards and ASIL level calculations are accurate |
| | Whether FTTI calculations match the scenario descriptions and are accurate | Confirm FTTI calculations match scenario descriptions and are accurate |
| HARA Analysis Details | Whether all items are analyzed to ensure no omissions | Confirm all items are analyzed without omissions |
| | Whether failure modes are fully analyzed through HAZOP | Confirm HAZOP analysis covers all failure modes |
| | Whether the scenario descriptions are concise, clear, and comprehensive, including key scenario elements (e.g., understandable by non-experts), and whether the scenario elements are fully covered (e.g., different road conditions, lighting conditions, etc.) | Confirm scenario descriptions are concise, clear, and cover all key elements |
| | Whether the sources of S, E, C ratings comply with regulations, whether different levels are distinguished, and whether E distinguishes between frequency and duration, with distinctions based on regulatory requirements | Confirm rating sources comply with regulations and distinctions are clear and based on regulatory requirements |
| | Whether similar safety goals are merged, and if so, whether the merged ASIL is set to the highest level, and whether FTTI is set to the shortest time | Confirm merged safety goals are set to the highest ASIL level and FTTI is set to the shortest time |
| | Whether the S, E, C ratings in HARA analysis meet the standards, and whether the ASIL level calculation meets the SEC combination results | Confirm ratings meet standards and ASIL level calculations are accurate |
| | Whether the formulated safety goals avoid corresponding failures | Confirm safety goals prevent failures |
| | Whether FTTI calculations match the scenario descriptions and are accurate | Confirm FTTI calculations match scenario descriptions and are accurate |
| FTA Analysis | Whether event decomposition is comprehensive, including self-failure, link failure, power supply failure, etc. | Confirm comprehensive event decomposition covering all failure modes |
| FSR Analysis | Whether there is a traceability relationship with SG, and whether the traced SG's ASIL level is consistent or meets ASIL decomposition requirements. Each FSR should have at least one corresponding SG. If an FSR has multiple SG traceability relationships, the ASIL level of the FSR should be set to the highest level among the multiple SGs | Confirm traceability relationship with SG and consistent or compliant ASIL levels |
| | Whether FSR attributes are complete, including requirement description, ID, safety state, ASIL level, FTTI, and deployed system | Confirm complete FSR attributes covering all necessary information |
| FSC Analysis | Whether the safety mechanism design can detect the fault, and whether the response includes a description of the safety state after detecting the fault | Confirm safety mechanisms can detect faults and responses include safety state descriptions |
| | Whether each FSR has a unique ID | Confirm each FSR has a unique ID |
| | Whether each fault in the FTA is covered by a corresponding FSR | Confirm each fault is covered by a corresponding FSR |
| | Whether each FSR has a corresponding time constraint and whether the formulation principles are reasonable | Confirm reasonable time constraints for each FSR |
| | Whether FSR descriptions clearly highlight the subsystem they relate to | Confirm clear FSR descriptions highlighting the subsystem |
| | Whether the FSR formulation avoids unreasonable arbitration under multiple functional requests and complies with regulatory requirements | Confirm reasonable FSR formulation without unreasonable arbitration, complying with regulatory requirements |

## D.2 Evaluation Criteria for Test cases

| Clause Number | Requirements |
|---|---|
| 1 | Each FSR should have at least one corresponding requirement |
| 2 | Each test case should include requirements for test methods and test case derivation methods |
| 3 | The selection of test methods should meet the ASIL level requirements of the associated FSR, with "++" indicating mandatory inclusion |
| 4 | The selection of test case derivation methods should meet the ASIL level requirements of the associated FSR, with "++" indicating mandatory inclusion |
| 5 | Test descriptions should be clear and unambiguous, test steps should be measurable, and expected test results should include signal names |
| 6 | The types of injected faults should meet all failure modes analyzed in the FTA |